\documentclass[usenatbib]{mnras}
\usepackage[T1]{fontenc}
\usepackage{ae,aecompl}

\usepackage{graphicx}	
\usepackage{amsmath}	
\usepackage{amssymb}	

\title[Photoionisation of S-like Chlorine]{Photoionisation of Cl$^+$ from the $3s^23p^4\;^3P_{2,1,0}$ and the
                                                               $3s^23p^4\;^1D_2, ^1S_0$ states in the  energy range 19 - 28 eV}
\author[B. M. McLaughlin]{Brendan M. McLaughlin,$^{1,2}$\thanks{E-mail:bmclaughlin899@btinternet.com}
\\
$^{1}$Center for Theoretical Atomic Molecular and Optical Physics (CTAMOP), 
            School of Mathematics and Physics, \\The David Bates Building,
            7 College Park,  Queens University Belfast, Belfast BT7 1NN, UK,\\
$^{2}$Institute for Theoretical Atomic Molecular and Optical Physics (ITAMP), 
           Harvard-Smithsonian Center for Astrophysics, \\
           60 Garden Street,  MS-14, Cambridge MA 02138, USA
}

\date{Accepted --\today ;  Revised \today; in original form 15 February 2016}

\pubyear{2016}

\begin{document}
\label{firstpage}
\pagerange{\pageref{firstpage}--\pageref{lastpage}}
\maketitle

\begin{abstract}
{Absolute photoionisation cross sections for the 
Cl$^+$ ion in its ground
and the metastable states; 
$3s^2 3p^4\; ^3P_{2,1,0}$, and $3s^2 3p^4\; ^1D_2,\; ^1S_0$, 
were measured recently at the Advanced Light Source 
(ALS) at Lawrence Berkeley National 
Laboratory using the merged beams photon-ion 
technique at an photon energy resolution of 15
meV  in the energy range 19  -- 28 eV.   
These measurements are compared with large-scale Dirac Coulomb
{\it R}-matrix calculations in the same energy range. 
Photoionisation of this sulphur-like chlorine ion
is characterized by multiple Rydberg series 
of autoionizing resonances superimposed on a direct
photoionisation continuum. A wealth of resonance 
features observed in the experimental spectra are
spectroscopically assigned and their 
resonance parameters tabulated and compared with the recent
measurements.  Metastable fractions in the parent ion beam
are determined from the present study.
Theoretical resonance energies and quantum defects of the 
prominent Rydberg series $3s^2 3p^3 nd$,  
identified in the spectra as $3p \rightarrow nd$ transitions
are compared with the available measurements made on this element. 
Weaker Rydberg series $3s^2 3p^3 ns$, 
identified as $3p \rightarrow ns$ transitions
and window resonances $3s3p^4 (^4P)np$ features, 
due to $3s \rightarrow np$ transitions 
are also found in the spectra.}

\end{abstract}



\section{Introduction}
In astrophysics, abundances calculations are  
based on available atomic data that often are 
 insufficient to make definite identifications of spectroscopic
 lines \citep{Cardelli1993}.  In planetary nebulae, 
 the known emission lines are used to identify a characteristic
 element resulting from the process of nucleosynthesis
 in stars \citep{sharpee2007,sterling2007}.  
 The study of photoionisation of sulphur-like ions is of
 considerable interest because of its abundance in space and the interstellar 
medium.  As previously indicated by Hern\'{a}ndez and co-workers \citep{Hernandez2015} there 
is a wealth of astrophysical applications for this Cl II ion 
\citep{Cartlidge2012,Sylwester2011,Neufeld2012,Moomey2012}.  
Sulphur-like chemistry is also of importance in theoretical studies
 of interstellar shocks \citep{Pineau1986}.  Photoabsorption and 
photoionisation processess in the vacuum ultraviolet (VUV) 
region play an important role in determining solar 
and stellar opacities \citep{Kohl1973,Dupree1978,Lombardi1981}.
The Cl II ion is an important basis for atmospheric 
and astrophysical models. Determining accurate abundances 
for sulphur-like chlorine is of great importance in
 understanding extragalactic H II regions \citep{Garnett1989} 
and emission lines of Cl II in the optical spectra of
 planetary nebulae NGC 6741 and IC 5117 \citep{Keenan2003}.
Cl II emission lines have also been seen 
in the spectra of the Io torus \citep{Schneider2000} 
and by the far-ultraviolet spectroscopic 
explorer (FUSE) \citep{Feldman2001}.

In the present study, we use a fully relativistic 
Dirac Coulomb  $R$-matrix approximation 
 \citep{norrington1987,norrington1991,norrington2004,grant2007}
to interpret and analyze resonance features found in recent high resolution 
measurements obtained at the ALS  \citep{Hernandez2015}.  In our work
we give a detailed interpretation and understanding of the atomic 
processes involved for single-photon ionization of Cl II forming Cl III.
The interaction of a single photon with a Cl II ion comprises contributions 
from both direct ionization and excitation of autoionizing resonances  \citep{berko1979}. 
Direct electron ejection processes relevant to the total cross section for 
single ionization of the Cl$^{+}$ ion in its ground configuration include,
\begin{equation}
h\nu + {\rm Cl}^{+}({3s^2 3p^4\,\,{^3P} }) \rightarrow \left\{ \begin{array} {l}
 {\rm Cl}^{2+}({3s^2 3p^3 }) + e^- \\
 {\rm Cl}^{2+}({3s 3p^4 }) + e^- .
\end{array} \right.
\end{equation}
Indirect ionization of Cl$^{+}$ levels within the $^3P$ ground term may proceed via resonance formation
\begin{equation}
h\nu + {\rm Cl}^{+}({3s^2 3p^4}\,\, {^3P}_J) \rightarrow \left\{ \begin{array} {l}
 {\rm Cl}^{+}({3s^2 3p^3} n\ell \,\, {^3L}_{J' }) \\
 {\rm Cl}^{+}({3s 3p^4} n\ell' \,\, {^3L}_{J' })
\end{array} \right.
\end{equation}
(where, $n\ell$= $ns$ or $nd$, and $n\ell'$=$np$), with subsequent decay by emission of a single electron
\begin{equation}
 {\rm Cl}^{+}({^3L}_{J' }) \rightarrow {\rm Cl}^{2+} + e^-,
\end{equation}
where $L$ is the total orbital momentum quantum number and $J'$ 
the total angular momentum quantum number 
of the intermediate resonant state. Selection rules for electric
 dipole transitions require that $J' = J$ or $J'=J\pm1$. 
Similar types of atomic processes occur for the case where this
 Cl$^+$ ion is in the excited metastable states,
 $3s^2 3p^4\,\,^1D_2,{\rm or}\, ^1S_0$.

On the experimental side, measurements of the photoionisation 
spectrum for this  ion have been reported on recently \citep{Hernandez2015}. 
Detailed high-resolution measurements were carried out at an energy 
resolution of 15 meV full width half maximum (FWHM) using synchrotron radiation at
the Advanced Light Source (ALS) in Berkeley, 
California, for the photon energies in the region 19 - 27.8 eV.
Absolute values for the cross section were reported on for
the photoionisation of the  $^3P_{2,1,0}$ states of the 
$3s^2 3p^4\; ^3P$ configuration in this sulphur-like ion, for photon energies in
the energy region 23.38 - 27.8 eV, and for the metastable
states $3s^2 3p^4\; ^1D_2,\; ^1S_0$ for photon energies, 
19.5 - 27.8 eV. Resonance features observed in the 
corresponding experimental spectra were analyzed and discussed 
but no attempt was made to assign and identify the resonance series \citep{Hernandez2015}.

On the theoretical side, previous photoionisation cross-section calculations of this ion 
to the author's knowledge have been rather limited \citep{Opacity1995}. 
We note that accurate transition probabilities  
\citep{fischer1977,cowan1981,fischer1997} 
and $f$-values between levels of a system
are good indicators of the quality of target wavefunctions  
used in subsequent cross sections calculations. 
For electron impact excitation (EIE) of the Cl$^{2+}$ (Cl III)
ion, calculations have been reported on using the intermediate 
coupling approximation by Sossah and Tayal  \citep{Sossah2012}
within the {\it R}-matrix method  
\citep{Burke1975,Burke2011} although no resonance features
were reported. Radiative data for chlorine and its ions
has also been reported on by Berrington and Nakazaki 
 \citep{Berrington2002} based on {\it R}-matrix calculations performed in 
$LS$-coupling as no data currently exists
 in the Opacity Project database  \citep{Opacity1995}. 

The limited high quality data available 
for the Cl$^+$ ion is one of the 
major motivating factors for the present investigations. 
Identification of the Rydberg  resonance 
series in spectra, in particular the window resonances 
is  another major motivation factor.  Understanding 
the interplay and  interaction  between the direct
 and indirect photoionisation processes help us 
understand and interpret the underlying physics.  
Furthermore, benchmarking
the present large-scale cross section calculations on this
 ion against available high-resolution experimental 
results, is essential as it provides further confidence in the data for use
in various laboratory and astrophysical plasma applications.

The layout of this paper is as follows. In
Section 2 we outline the theoretical methods employed
in our work. Section 3 presents the theoretical results 
from the DARC photoionisation cross and the resonance analysis.  
Section 4 presents a comparison between the available experimental 
measurements \citep{Hernandez2015} 
and the theoretical cross section results 
for singly ionized atomic chlorine ground 
and metastable terms [Cl$^+$ ($3s^2 3p^4\; ^3P_{2,1,0}$ )
and Cl$^+$ ($3s^2 3p^4 \; ^1D_2,\; ^1S_0$ )] 
in the photon energy range from 19 - 28 eV. 
Section 5 gives a brief discussion 
of our results in comparison to the available 
measurements \citep{Hernandez2015}. 
Finally in Section 6 we give a summary
of our findings from our theoretical work.

		%
		%
\begin{figure*}
\centering
\includegraphics[width=\textwidth]{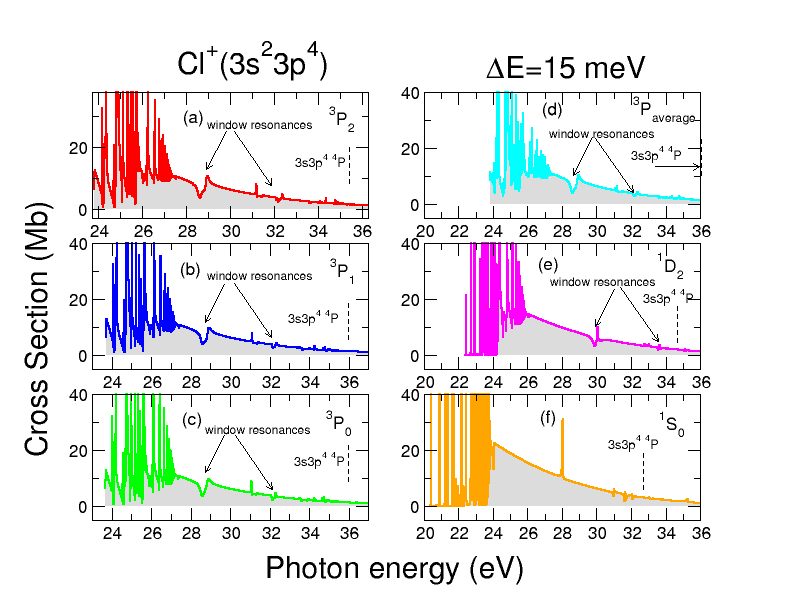}
\caption{\label{all} (colour online) Theoretical cross sections from the 512 level DARC calculations for the sulphur-like chlorine ion in the
		$3s^23p^4\; ^3P_{2,1,0}$, $3s^23p^4\; ^1D_2$, and $3s^23p^4\; ^1S_0$ initial states 
		convoluted with a Gaussian profile of 15 meV. Single photoionisation
		cross sections of the sulphur-like chlorine ion as a function of energy over the photon energy from thresholds to and energy region
		just above the Cl$^{2+}$($3s3p^4\; ^4P$) threshold, illustrating strong resonance features in the spectra. 
		(a) $3s^23p^4\; ^3P_2$, (b) $3s^23p^4\; ^3P_1$, (c) $3s^23p^4\; ^3P_0$, (d) $3s^23p^4\; ^3P$ level averaged, 
		(e) metastable $3s^23p^4\; ^1D_2$ and (f) metastable $3s^23p^4\; ^1S_0$ cross sections. The
		corresponding series limits $E_{\infty}$ of Equation (2) for the window resonances 
		converging to the $3s3p^4\; ^4P$ threshold is indicated by
		vertical lines. }
\end{figure*}
\section{Theory}

\subsection{Dirac Coulomb {\it R}-matrix}

The study of the photoabsorption spectrum of Mid-Z  elements and 
their isoelectronic sequences is very interesting due to the 
open-shell nature of these complexes and the role 
played by electron correlation effects. 
Singly ionized atomic chlorine is one such open-shell ion. 
In a similar manner to our previous work on atomic sulphur \citep{Berlin2015}, to gauge the 
quality of our work we carried out 
large-scale close-coupling calculations and benchmarked the results
with the experimental photoionisation 
cross section measurements  \citep{Hernandez2015}.  
 The target wavefunctions for our work were obtained using the GRASP code  
\citep{dyall1989,grant2006,grant2007}, and the subsequent
 theoretical photoionisation  cross sections 
 were obtained with the DARC codes \citep{darc}. 
All the target wavefunction orbitals up 
to $n$ = 3 were determined within an 
EAL calculation using the GRASP code.
Target wavefunctions were then generated 
with all 512 levels arising from the eight configurations: 
 $3s^23p^3$,  $3s3p^4$,  $3s^23p^23d$, $3s^23p3d^2$,
 $3s3p^33d$,  $3p^33d^2$, $3p^5$, $3s3p^23d^2$, of the
 residual sulphur singly ionized ion. All these levels were included  
 in the close-coupling expansion. 
 Our photoionisation cross section calculations were 
performed within the relativistic 
Dirac-Coulomb {\it R}-matrix approximation \citep{darc,grant2007}. 
An efficient parallel version of the 
DARC suite of codes has been developed 
to address the challenge of electron 
and photon interactions with atomic systems catering for hundreds of 
levels and thousands of scattering channels. 

For comparison with high-resolution measurements 
made at the ALS, state-of-the-art theoretical methods 
with highly correlated wavefunctions that 
include relativistic effects are used.
Due to the presence of metastable states in the Cl$^{+}$ ion beam 
experiments additional theoretical calculations were
required in order to have a true comparison between theory and experiment. 
Recent modifications to the Dirac-Atomic {\it R}-matrix-Codes (DARC) 
 \citep{darc,ballance2006,Ballance2012,
 McLaughlin2012,McLaughlin2015a,McLaughlin2015b,McLaughlin2016a}  
now allow high quality photoionisation cross section 
calculations to be made (in a timely manner) 
on heavy complex systems (Fe-peak elements and mid-Z atoms) of prime 
interest to astrophysics and plasma applications. 
Cross-section calculations for various trans-Fe element single 
photoionisation of Se$^+$ \citep{Ballance2012}, 
Se$^{2+}$ \citep{david2015},
 Xe$^+$ \citep{McLaughlin2012}, 
 Kr$^+$ \citep{Hino2012}, Xe$^{7+}$ \citep{Mueller2014},  
$2p^{-1}$ inner-shell studies on 
Si$^+$ ions  \citep{Kennedy2014}, valence-shell studies on 
neutral sulphur \citep{Berlin2015}, Tunsgten and its
 ions  \citep{Ballance2015,Mueller2015, McLaughlin2016}
 have been made using these DARC codes.  
 We point out that suitable agreement of DARC 
photoionisation cross-sections calculations with the 
above high resolution measurements made at leading 
 synchrotron light sources such as 
 ALS, ASTRID, SOLEIL and PETRA III has been obtained.

		%
		%

\begin{figure*}
\centering
\includegraphics[width=\textwidth]{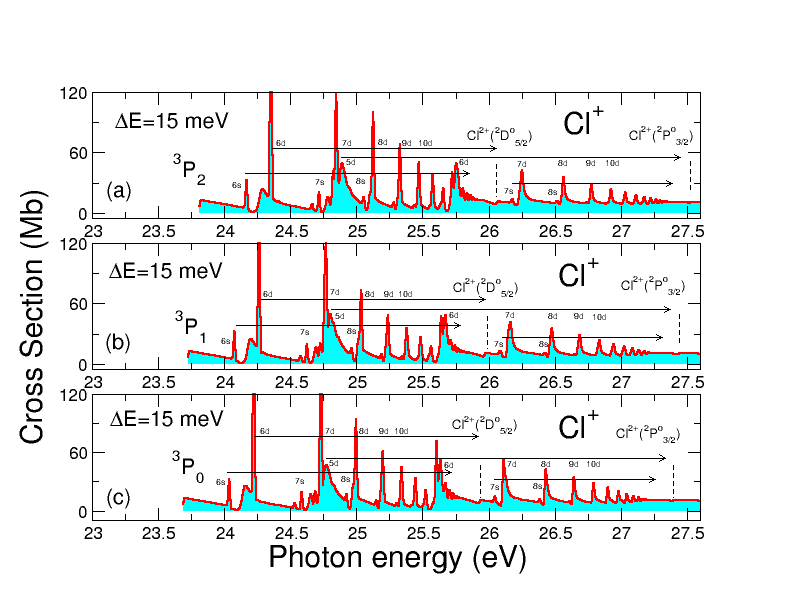}
\caption{\label{thres-3P}(colour online) Single photoionisation of the sulphur-like 
		chlorine ion as a function of the photon energy from the
		$3s^23p^4\; ^3P_{2,1,0}$ states from threshold to 27.6 eV. 
		(a) $3s^23p^4\; ^3P_2$, (b) $3s^23p^4\; ^3P_1$, and (c) $3s^23p^4\; ^3P_0$.  
		Strong $3s^23p^3(^2D^o_{5/2 })nd$ Rydberg series are present as are the much weaker 
		$3s^23p^3(^2D^o_{5/2 })ns$ series converging to the Cl$^+$($^2D^o_{5/2}$) threshold. 
		Similar strong $3s^23p^3(^2P^o_{3/2 })nd$ 
		Rydberg series are present as are the much weaker 
		$3s^23p^3(^2P^o_{3/2 })ns$ series converging to the Cl$^+$($^2P^o_{3/2}$) threshold.
		Theoretical cross section calculations were carried out with the DARC codes,
		and convoluted with a Gaussian having a profile of 15 meV. 
		The assigned Rydberg series are indicated as vertical lines grouped
		by horizontal lines. The corresponding series limits $E_{\infty}$ of Equation (2) 
		for each series are indicated by a vertical
		lines at the end of the line groups. Several of the $nd$ interloping resonances 
		are seen to disrupt the regular Rydberg resonance pattern of the spectra. 
		Resonance energies and quantum defects for the various series are tabulated in Tables \ref{res-tab1} and
		\ref{res-tab2} and compared with the available experimental values.}
\end{figure*}

\subsection{Photoionisation}

To benchmark theory with the recent ALS measurements of
photoionisation cross-sections \citep{Hernandez2015}, 
calculations on this sulphur-like ion were 
performed on the ground and the excited metastable levels 
associated with the $3s^23p^4$ configuration. 
Hibbert and co-workers have shown that two-electron 
promotions are important to include to get accurate 
energies, $f$-values and 
Einstein coefficients  \citep{Ohja1987,Keenan1993} which 
are included in the present study.
In our photoionisation cross-section calculations for 
this element, all 512 levels arising from the eight configurations: 
$3s^23p^3$, $3s3p^4$, $3s^23p^23d$, $3s^23p3d^2$,
$3s3p^33d$, $3p^33d^2$, $3p^5$, $3s3p^23d^2$ 
of the residual chlorine doubly ionized ion
were included in the close-coupling expansion.  
In our calculations for the ground and 
metastable levels, the outer region electron-ion collision 
problem was solved with an extremely fine energy  mesh of 0.272$\mu$eV  
to fully resolve the extremely narrow resonance 
features found in the appropriate photoionisation cross sections.  

Photoionisation cross section calculations with this 512-level model 
 were performed for the $^3P_{2}$ ground and
 the metastable $^3P_{1,0}$, $^1D_2$ and $^1S_{0}$ states of this ion, 
over the photon energy range similar to experimental studies.  
All the cross section calculations 
were carried out within the Dirac-Coulomb {\it R}-matrix approximation  
  \citep{darc,ballance2006,Ballance2012,McLaughlin2012,McLaughlin2015a,McLaughlin2015b}.

The $^3P_{2,1,0}$ levels require the bound-free dipole 
 matrices, J$^{\pi}$ = 2$^e$,1$^e$, 0$^e$ \ $\rightarrow$ \ J$^{{\prime}\pi^{\prime}}$ = 0$^o$,1$^o$,2$^o$,3$^o$ 
and for the excited $^1D_2$ and $^1S_0$ metastable states, 
 the bound-free dipole matrices, J$^{\pi}$ = 0$^e$, 2$^e$ \  $\rightarrow$ 
J$^{{\prime}\pi^{\prime}}$= 1$^o$, 2$^o$, 3$^o$. 
It was necessary to carry out these 
additional photoionisation cross-section calculations 
 to span the entire photon energy range  of 
 the experimental measurements, where 
 various metastable states are present in the beam 
 in order to have a true comparison between theory and experiment. 
 The $jj$ -coupled Hamiltonian diagonal matrices 
were adjusted so that the theoretical term energies matched 
the recommended experimental values of the NIST tabulations  \citep{NIST2014}. 
We note that this energy adjustment 
ensures better positioning of resonances relative to all thresholds included 
in the present calculations.

	%
%
	%
	%
\begin{table*}
\caption{\label{res-tab1} Principal quantum numbers $n$, resonance energies (eV), and quantum defects  $\mu$
	of the Cl$^+$($3s^23p^3[^2D^o_{5/2}])ns, nd$ Rydberg series seen in the Cl$^+$($3s^23p^4\; ^3P_{2,1,0}$) 
	photoionisation spectra, converging to the Cl$^+$($3s^23p^3[^2D^o_{5/2}]$) threshold. The
	experimental resonance energies  \citep{Hernandez2015} are calibrated to $\pm$13 meV. 
	The theoretical results are obtained from the 512-level 
	DARC calculations performed within the Dirac Coulomb {\it R}-matrix approximation. The
	experimental entries in parenthesis are uncertain.}
\begin{tabular}{cccccccccc}
\hline\hline
  Cl$^+$	& 		&$E_n (eV)$	&$E_n (eV)$	&$\mu$	 &$\mu$		&$E_n (eV)$ 	&$E_n (eV)$	&$\mu$		&$\mu$\\
(Initial State)	& 		&Expt$^a$	&Theory$^b$	&Expt$^a$&Theory$^b$	&Expt$^a$	&Theory$^b$	&Expt$^a$	&Theory$^b$ \\
\hline
		&$n$ 		&	&$3s^23p^3(^2D^o_{5/2 })nd$	&	&	&&$3s^23p^3(^2D^o_{5/2 })ns$ 	&		&\\	
$3s^23p^4\;^3P_2$ &		&		&			&		&		&		&		&		&\\
		&6		&24.348		&24.353		&0.38		&0.35		&		&24.166	&		&0.64\\
		&7		&24.829		&24.846		&0.35		&0.30		&		&24.710	&		&0.64\\	
		&8		&25.128		&25.130		&0.36		&0.35		&		&25.056	&		&0.64\\
		&9		&25.335		&25.334		&0.34		&0.34		&		&25.283	&		&0.63\\
		&10		&25.479		&25.476		&0.32		&0.35		&		&		&		&\\	
		&11		&25.583		&25.584		&0.32		&0.30		&		&		&		&\\
		&$\dots$	&			&			&		&		&		&		&		&\\
		&$\infty$	&26.060$^c$	&26.060$^c$	&		&		&26.060$^c$&26.060$^c$&		&\\
\\
		&$n$ 		&	&$3s^23p^3(^2D^o_{5/2 })nd$	&	&	& 	&$3s^23p^3(^2D^o_{5/2 })ns$ 	&		&\\	
$3s^23p^4\;^3P_1$ &		&		&		&	&		&		&		&		&\\
		&6		&24.259		&24.264		&0.37		&0.36		&		&24.075	&		&0.65\\
		&7		&24.750		&24.762		&0.33		&0.30		&		&24.630	&		&0.64\\
		&8		&25.036		&25.039		&0.38		&0.37		&		&		&		&\\
		&9		&25.238		&25.237		&0.40		&0.40		&		&		&		&\\
		&10		&25.384		&25.385		&0.42		&0.40		&		&		&		&\\
		&11		&(25.492)		&25.493		&(0.37)	&0.36		&		&		&		&\\
		&$\dots$	&			&			&		&		&		&		&		&\\
		&$\infty$	&25.974$^c$	&25.974$^c$	&		&		&25.974$^c$&25.974$^c$&		&\\
\\
		&$n$ 		&		&$3s^23p^3(^2D^o_{5/2 })nd$	&	&		&	&$3s^23p^3(^2D^o_{5/2 })ns$ &	&\\	
$3s^23p^4\;^3P_0$ &	&			&			&		&		&		&		&		&\\
		&6		&24.219		&24.223		&0.37		&0.36		&		&24.041	&		&0.64\\
		&7		&			&24.733		&		&0.27		&		&24.591	&		&0.64\\
		&8		&(25.000)		&24.999		&(0.38)	&0.38		&		&24.926	&		&0.64\\
		&9		&			&25.198		&		&0.40		&		&		&		&\\
		&10		&			&25.345		&		&0.40		&		&		&		&\\
		&$\dots$	&			&			&		&		&		&		&		&\\
		&$\infty$	&25.936$^c$	&25.936$^c$	&		&		&25.936$^c$&25.936$^c$&		&\\		
\hline\hline
\end{tabular}
\begin{flushleft}
$^a$ALS experimental results  \citep{Hernandez2015}.\\
$^b$Present {\it R}-matrix DARC calculations.\\
$^c$Rydberg series limits $E_{\infty}$ for Cl$^+$ are from the  NIST tabulations  \citep{NIST2014}.
\end{flushleft}
\end{table*}

	%
	%
\begin{table*}
\caption{\label{res-tab2} Principal quantum numbers $n$, resonance energies (eV), and quantum defects  $\mu$
	of the Cl$^+$($3s^23p^3[^2P^o_{3/2}])ns, nd$ Rydberg series seen in the Cl$^+$($3s^23p^4\; ^3P_{2,1,0}$) 
	photoionisation spectra, converging to the Cl$^+$($3s^23p^3[^2P^o_{3/2}]$) threshold. The
	experimental resonance energies  \citep{Hernandez2015} are calibrated to $\pm$13 meV. 
	 The theoretical results are obtained from the 512-level 
	DARC calculations performed within the Dirac Coulomb {\it R}-matrix approximation. The
	experimental entries in parenthesis are uncertain.}
\begin{tabular}{cccccccccc}
\hline\hline
  Cl$^+$	& 		&$E_n (eV)$	&$E_n (eV)$	&$\mu$	 &$\mu$		&$E_n (eV)$ 	&$E_n (eV)$	&$\mu$		&$\mu$\\
(Initial State)	& 		&Expt$^a$	&Theory$^b$	&Expt$^a$&Theory$^b$	&Expt$^a$	&Theory$^b$	&Expt$^a$	&Theory$^b$ \\
\hline
		&$n$ 		&			&$3s^23p^3(^2P^o_{3/2})nd$	&	&&&$3s^23p^3(^2P^o_{3/2})ns$ 	&		&\\	
$3s^23p^4\;^3P_2$ &	&			&			&			&		&		&		&		&\\
		&5		&(24.869)		&24.880		&(0.47)		&0.46		&		&		&		&\\
		&6		&(25.745)		&25.754		&(0.47)		&0.45		&		&		&		&\\
		&7		&26.246		&26.253		&0.47			&0.45		&		&26.180	&		&0.63\\	
		&8		&26.564		&26.576		&0.46			&0.46		&		&26.519	&		&0.63\\
		&9		&26.778		&26.780		&0.45			&0.44		&		&		&		&\\
		&10		&26.928		&26.927		&0.43			&0.44		&		&		&		&\\	
		&11		&27.031		&27.035		&0.47			&0.43		&		&		&		&\\
		&12		&(27.114)		&27.115		&(0.45)		&0.44		&		&		&		&\\
		&13		&27.175		&27.178		&0.48			&0.42		&		&		&		&\\
		&$\dots$	&			&			&			&		&		&		&		&\\
		&$\infty$	&27.522$^c$	&27.522$^c$	&			&		&27.522$^c$&27.522$^c$&		&\\
\\
		&$n$ 		&	&$3s^23p^3(^2P^o_{3/2 })nd$	&	&		& &$3s^23p^3(^2P^o_{3/2 })ns$ 	&		&\\	
$3s^23p^4\;^3P_1$ &		&		&			&			&		&		&		&		&\\
		&5		&(24.783)		&24.804		&(0.47)		&0.45		&		&		&		&\\
		&6		&(25.660)		&25.668		&(0.46)		&0.45		&		&		&		&\\
		&7		&26.157		&26.162		&0.47			&0.46		&		&26.088	&		&0.64\\
		&8		&26.475		&26.480		&0.47			&0.45		&		&26.429	&		&0.64\\
		&9		&26.687		&26.695		&0.46			&0.43		&		&		&		&\\
		&10		&26.833		&26.842		&0.49			&0.42		&		&		&		&\\
		&$\dots$	&			&			&			&		&		&		&		&\\
		&$\infty$	&27.435$^c$	&27.435$^c$	&			&		&27.435$^c$&27.435$^c$&		&\\
\\
		&$n$ 		&			&$3s^23p^3(^2P^o_{3/2 })nd$	&	&		&	&$3s^23p^3(^2P^o_{3/2 })ns$ &  &\\	
$3s^23p^4\;^3P_0$ &	&			&			&			&		&		&		&		&\\
		&5		&(24.714)		&24.767		&(0.50)		&0.45		&		&		&		&\\
		&6		&			&25.635		&			&0.44		&		&26.048	&		&0.65\\
		&7		&26.108		&26.117		&0.51			&0.48		&		&26.383	&		&0.68\\
		&8		&26.430		&26.434		&0.50			&0.49		&		&		&		&\\
		&9		&26.648		&26.650		&0.49			&0.47		&		&		&		&\\
		&10		&			&26.797		&			&0.48		&		&		&		&\\
		&$\dots$	&			&			&			&		&		&		&		&\\
		&$\infty$	&27.398$^c$	&27.398$^c$	&			&		&27.398$^c$&27.398$^c$&		&\\		
\hline\hline
\end{tabular}
\begin{flushleft}
$^a$ALS experimental results  \citep{Hernandez2015}.\\
$^b$Present {\it R}-matrix DARC calculations.\\
$^c$Rydberg series limits $E_{\infty}$ for Cl$^+$ are from the  NIST tabulations  \citep{NIST2014}.
\end{flushleft}
\end{table*}
		%
		%

\begin{figure*}
\centering
\includegraphics[width=\textwidth]{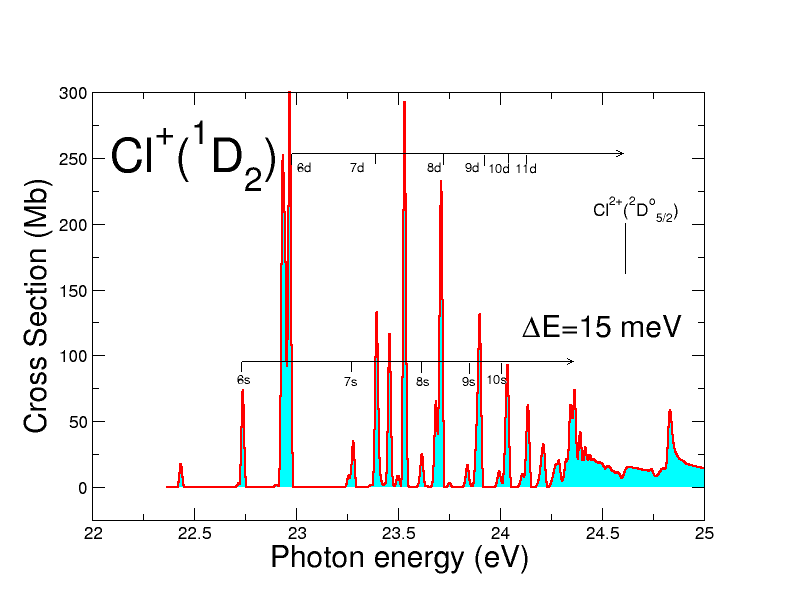}
\caption{\label{thres-1D}(colour online) Single photoionisation of the sulphur-like chlorine ion 
			as a function of the photon energy in the $3s^23p^4\; ^1D_2$
			metastable state from threshold to 25 eV. Theoretical cross section were carried out with the DARC codes, and convoluted
			with a Gaussian having a profile of 15 meV. The assigned Rydberg series are indicated as vertical lines grouped by horizontal
			 lines. The corresponding series limits $E_{\infty}$ 
			of Equation (2) for each series are indicated by a vertical line
			at the end of the line groups. The strong $3s^23p^3(^2D^o_{5/2 })nd$ 
			Rydberg series are present as are the much weaker 
			$3s^23p^3(^2D^o_{5/2 })ns$ series converging to the Cl$^+$($^2D^o_{5/2}$) threshold. Several of the $nd$
			 interloping resonances are seen to disrupt the regular Rydberg resonance pattern of
			the spectra. The first few values of $n$ for each series is displayed 
			close to its corresponding vertical line in each group. Resonance
			energies and quantum defects for the various series are tabulated in Table \ref{res-tab3} and compared with the available experimental
			values.}
\end{figure*}
	%
	%
\begin{table*}
\caption{\label{res-tab3} Principal quantum numbers $n$, resonance energies (eV), and quantum defects  $\mu$
	of the Cl$^+$($3s^23p^3[^2P^o_{1/2}])ns, nd$ and Cl$^+$($3s^23p^3[^2D^o_{5/2}])ns, nd$ 
	Rydberg series seen in the Cl$^+$($3s^23p^4\; ^1D_2~ {\rm and}~ ^1S_0$) 
	photoionisation spectra, converging to the Cl$^+$($3s^23p^3[^2P^o_{1/2}]$) and Cl$^+$($3s^23p^3[^2P^o_{1/2}]$) thresholds. 
	 The theoretical results are obtained from the 512-level 
	DARC calculations performed within the Dirac Coulomb {\it R}-matrix approximation.
	The experimental values for the resonance energies  \citep{Hernandez2015} 
	are from the recent ALS measurements calibrated to $\pm$13 meV uncertainity.
	The experimental entries in parenthesis are uncertain.}
\begin{tabular}{cccccccccc}
\hline\hline
  Cl$^+$	& 		&$E_n (eV)$	&$E_n (eV)$	&$\mu$	 &$\mu$		&$E_n (eV)$ 	&$E_n (eV)$	&$\mu$		&$\mu$\\
(Initial State)	& 		&Expt$^a$	&Theory$^b$	&Expt$^a$&Theory$^b$	&Expt$^a$	&Theory$^b$	&Expt$^a$	&Theory$^b$ \\
\hline
		&$n$ 		&			&$3s^23p^3(^2P^o_{1/2 })nd$	&	&&&$3s^23p^3(^2P^o_{1/2 })ns$ 	&		&\\	
$3s^23p^4\;^1S_0$ &	&			&			&	&		&		&			&		&\\
		&4		&20.426		&20.420		&0.13	&0.13		&		&			&		&\\
		&5		&21.742		&21.734		&0.15	&0.16		&		&21.235		&		&0.61\\
		&6		&22.463		&22.459		&0.15	&0.16		&		&22.191		&		&0.60\\
		&7		&22.900		&22.890		&0.13	&0.16		&		&22.724		&		&0.60\\	
		&8		&23.178		&23.170		&0.12	&0.15		&		&23.061		&		&0.60\\
		&9		&			&23.361		&	&0.14		&		&23.282		&		&0.60\\
		&$\dots$	&			&			&	&		&		&			&		&\\
		&$\infty$	&24.054$^c$	&24.054$^c$	&	&		&24.054$^c$&24.054$^c$	&		&\\
\\
		&$n$ 		&			&$3s^23p^3(^2D^o_{5/2 })nd$&	&& &$3s^23p^3(^2D^o_{5/2 })ns$ &	&\\	
$3s^23p^4\;^1D_2$ &	&			&			&	&		&		&			&		&\\
		&6		&22.979		&22.969		&0.27	&0.25		&		&22.741		&		&0.61\\
		&7		&			&23.392		&	&0.32		&		&23.278		&		&0.62\\
		&8		&23.718		&23.719		&0.21	&0.21		&		&23.616		&		&0.62\\
		&9		&23.907		&23.907		&0.23	&0.23		&		&23.841		&		&0.62\\
		&10		&24.040		&24.039		&0.27	&0.28		&		&23.995		&		&0.63\\
		&11		&(24.152)		&24.138		&(0.27)		&0.32		&			&		&	\\
		&$\dots$	&			&			&			&		&			&		&	\\
		&$\infty$	&24.615$^c$	&24.615$^c$	&			&		&24.615$^c$	&24.615$^c$&	\\
\hline\hline
\end{tabular}
\begin{flushleft}
$^a$ALS experimental results  \citep{Hernandez2015}.\\
$^b$Present {\it R}-matrix DARC calculations.\\
$^c$Rydberg series limits $E_{\infty}$ for Cl$^+$ are from the  NIST tabulations  \citep{NIST2014}.
\end{flushleft}
\end{table*}
		%
		%

\begin{figure*}
\centering
\includegraphics[width=\textwidth]{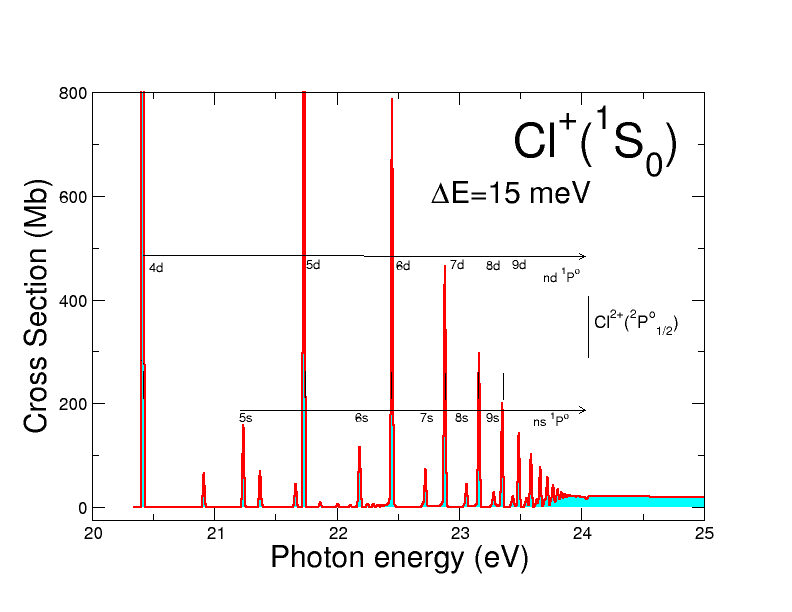}
\caption{\label{thres-1S}(colour online) Single photoionisation of the sulphur-like chlorine ion as a function of the photon energy in the $3s^23p^4\; ^1S_0$
			metastable state from threshold to 25 eV. Theoretical cross section calculations were carried out with the DARC codes, and
			convoluted with a Gaussian having a profile of 15 meV. The assigned Rydberg series are indicated as vertical lines grouped by
			horizontal or inclined lines. The corresponding series limits $E_{\infty}$ of Equation (2) for each series are indicated by a vertical 
			lines at the end of the line groups. The strong $3s^23p^3(^2P^o_{1/2 })nd$ Rydberg series are present as are the much weaker
			$3s^23p^3(^2P^o_{1/2 })ns$ series converging to the Cl$^+$($^2P^o_{1/2}$) threshold. 
			Resonance energies and quantum defects for the various series are tabulated in Table \ref{res-tab3} and compared
			with the available experimental values}
\end{figure*}
		%
		%
\begin{figure*}
\centering
\includegraphics[width=\textwidth]{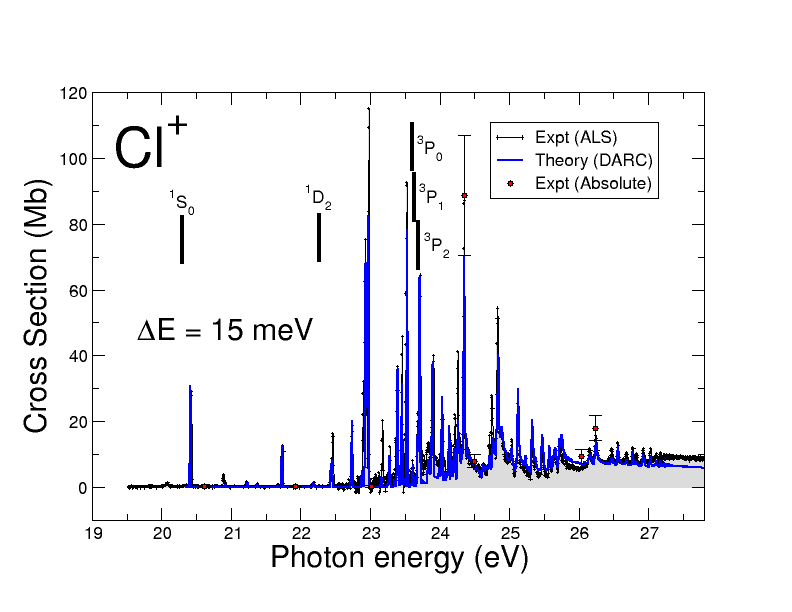}

\caption{\label{overview}(colour online) Single photoionisation of the sulphur-like chlorine ion as a function of the photon energy in the energy
			region 19 - 27.5 eV. The experimental results shown are from the recent high resolution measurements made at the ALS
			at a resolution of 15 meV FWHM \citep{Hernandez2015}. 
			The theoretical cross section calculations are from large-scale DARC calculations
			convoluted with a Gaussian having a profile of 15 meV FWHM. An appropriate weighting has been used for the ground and
			metastable states of this sulphur-like chlorine ion to best match the high resolution ALS experimental measurements (see text
			for details).}
\end{figure*}

\subsection{Resonances}

The energy levels available from the NIST tabulations  \citep{NIST2014} 
were used as a helpful guide for the present assignments.
The resonance series identification can be made from Rydberg's
 formula \citep{rydberg1,rydberg2,rydberg3,eisberg1985} :
\begin{equation}
	\epsilon_n= \epsilon_{\infty} - \frac{{\cal{Z}}^2}{\nu^2}
\end{equation}
where in Rydbergs $\epsilon_n$ is the transition 
energy, $\epsilon_{\infty}$ is the
ionization potential of the excited 
electron to the corresponding final state ($n$ = $\infty$),
 i.e. the resonance series limit and 
 $n$ being the principal quantum number.
The relationship between the principal quantum number
$n$, the effective quantum number $\nu$  and the quantum 
defect $\mu$ for an ion of effective charge $\cal{Z}$ is given by 
 $\nu$ = $n$ - $\mu$  \citep{Shore1967,Seaton1983}. 
 Converting all quantities to eV we 
 can represent the Rydberg series as;
\begin{equation}
	E_n= E_{\infty} - \frac{{\cal{Z}}^2 R}{(n - \mu)^2}
\end{equation}
Here, $E_n$ is the resonance energy, $E_{\infty}$ the resonance series
limit, $\cal{Z}$  is the charge of the core (in this case $\cal{Z}$ = 2), $\mu$
is the quantum defect, being zero for a pure hydrogenic
state.  For a hydrogenic system this can be written as,
\begin{equation}
	E^{\rm H}_n= E_{\infty} - \frac{{\cal{Z}}^2 R}{n^2}
\end{equation}
where the Rydberg constant $R$ is 13.605698 eV.

The multi-channel {\it R}-matrix eigenphase derivative
(QB) technique, which is applicable to atomic and molecular complexes, 
as developed by Berrington and co-workers  \citep{qb1,qb2,qb3} 
was used to locate and determine the resonance positions. 
The resonance width $\Gamma$ may also
be determined from the inverse of the energy derivative
of the eigenphase sum  at the resonance energy $E_r$ via
\begin{equation}
	\Gamma = 2\left[ \frac{d\delta}{dE}\right]^{-1}_{E=E_r} = 2\left[\delta^{\prime}\right]^{-1}_{E=E_r} .
\end{equation}
Finally, in order to compare with experimental, 
the theoretical cross-section calculations were
convoluted with a Gaussian having a profile 
of width similar to the ALS experiment resolution (15 meV FWHM).
Due to the presence of metastable states 
 in the Cl$^+$ ion beam experiments, 
a suitable admixture of the theoretical 
initial state populations was required. 
Details are outlined in the following sections. 

\section{Results}

\subsection{Cross Sections}

The electronic ground-state term of singly ionized
atomic chlorine in $LS$ coupling is $3s^23p^4\; ^3P$ and by spin-orbit 
effects splits into the three fine-stucture states namely,
$3s^23p^4\; ^3P_{2,1,0}$. The difference in energy between the
ground state $3s^23p^4\; ^3P_2$, 
the metastable $3s^23p^4\; ^3P_1$, and $3s^23p^4\; ^3P_0$
states is 86.3 and 123.5 meV  \citep{NIST2014}, respectively. 
Since additional metastable states, $3s^23p^4\; ^1S_0, \;^1D_2$ are 
present in the parent ion beam of the ALS experiments,
  cross-section calculations for these states were required to
span the energy region investigated. In order to simulate
the experimental conditions the theoretical cross section
are convoluted with a Gaussian having 
a 15 meV full width half maximum (FWHM)
profile and an appropriate fraction of all these excited states
are required to compare theory directly with experiment.

Fig. \ref{all} shows an overview of our results for all five initial
states of this sulphur-like ion convoluted with a Gaussian
of 15 meV FWHM from their respective thresholds up
to a photon energy just above the $3s3p^4\; ^4P$ state of the
doubly ionized chlorine ion. As seen from 
Fig. \ref{all} strong resonances features are found in the spectra
below about 27 eV. Above the Cl$^{2+}$($^2P^o_{3/2}$) 
threshold a prominent series of window resonances 
$3s3p^4(^4P)np$ converging to the 
Cl$^{2+}$($3s3p^4\; ^4P$) threshold is clearly visible
in the spectra. The origin of such window resonances
has already been discussed in our recent work on atomic
sulphur  \citep{Berlin2015}. Suffice it is to 
say that in atomic sulphur, the
$3s3p^4\; ^4P$ and $3s3p^4\; ^2P$ ionization states (occurring from
a $3s$ vacancy state) are located 
at 19.056 eV and 22.105 eV, respectively,
 and two prominent Rydberg series of
window resonances were found. We note that similar
types of structures were observed 
by Angel and Samson  \citep{Samson1988} 
in atomic oxygen, in atomic selenium, 
by Gibson and co-workers \citep{Gibson1986}, 
and atomic tellurium, by Berkowitz and co-workers 
 \citep{Berk1981}. A detailed discussion 
 of the various resonance features 
found in the respective theoretical spectra for this
sulphur-like chlorine ion is presented below and comparisons made
with the high resolution measurements recently made at
the ALS  \citep{Hernandez2015}.

\subsection{Photoionisation of the Cl$^+$($3s^23p^4$\;$^3P_{2,1,0}$) states}

Photoionisation of singly ionized atomic chlorine from
the initial $3s^23p^4\; ^3P_{2,1,0}$ fine-structure levels 
was investigated using large-scale {\it R}-matrix calculations 
within the Dirac Coulomb approximation (DARC). 
The electronic configuration of the
doubly ionized chlorine ion (Cl$^{2+}$) in the $3s^23p^3$ configuration 
forms the states $^4S^o_{3/2}$, $^2D^o_{3/2,5/2}$, and $^2P^o_{1/2, 3/2}$.
As illustrated in Fig. \ref{thres-3P}, the strongest Rydberg 
resonances series in the respective $3s^23p^4\; ^3P_{2,1,0}$ spectra 
converging to the Cl$^{2+}$($^2D^o_{5/2}$) 
and Cl$^{2+}$($^2P^o_{3/2}$) thresholds
are the $3s^23p^3(^2D^o_{5/2 })nd$ 
and $3s^23p^3(^2P^o_{3/2 })nd$ type series. 
Weaker Rydberg series $3s^23p^3(^2D^o_{5/2 })ns$
and $3s^23p^3(^2P^o_{3/2 })ns$ are also 
found in the respective theoretical spectra. 
Only the energies and quantum defects of 
the prominent $3s^23p^3nd$ series were tabulated
in the ALS experimental work \citep{Hernandez2015} but were not 
spectroscopically assigned. A detailed discussion 
of such Rydberg  series has already been given in our
recent study on atomic sulphur \citep{Berlin2015} and will not be 
expanded upon here. Suffice it is to say that the dominant 
series is the $3s^23p^3nd$ with the weaker series being the $3s^23p^3ns$. 
The energies and quantum defects are
tabulated in Table \ref{res-tab1} for both types of series converging
 to the Cl$^{2+}$($^2D^o_{5/2}$) threshold and in Table \ref{res-tab2}
 converging to Cl$^{2+}$($^2P^o_{3/2}$) threshold for all the respective
$3s^23p^4\; ^3P_{2,1,0}$ initial states. 
From Tables \ref{res-tab1} and \ref{res-tab2} 
excellent agreement of our theoretical values with the available
high resolution results from the ALS for the prominent
$3s^23p^3nd$ Rydberg series is seen. The majority of the
theoretical energies for the respective $3s^23p^3nd$ Rydberg
resonance series are within the $\pm$13 meV experimental
uncertainity. As illustrated in Fig. \ref{thres-3P}, the $5d$ and $6d$ 
members of the $3s^23p^3(^2P^o_{3/2 })nd$ series are interloping 
resonances, lying below the Cl$^{2+}$($^2D^o_{ 5/2}$) threshold and 
perturbing the spectrum of the $3s^23p^3(^2D^o_{5/2 })nd$ Rydberg
resonance series. Finally, we have also 
tabulated several members of the weaker
$3s^23p^3ns$ Rydberg series found in the theoretical spectrum 
in Tables \ref{res-tab1} and \ref{res-tab2} for completeness.	

\subsection{Photoionisation of the metastable Cl$^+$($3s^23p^4\;^1D_{2}$, and $^1S_0$) states}

Photoionisation cross section calculations of the
singly ionized atomic chlorine ion from the initial
$3s^23p^4\; ^1D_2$ and $^1S_0$ metastable states were also investigated 
using the same large-scale {\it R}-matrix model within
the Dirac Coulomb approximation (DARC), as these
states were present in the parent ion beam of the recent 
ALS measurements \citep{Hernandez2015}. Fig. \ref{thres-1D} shows the results
for the $3s^23p^4\; ^1D_2$ initial state and Fig. \ref{thres-1S} the 
corresponding $3s^23p^4\; ^1S_0$ case. Cross sections calculations are
shown for photon energies from their respective thresholds 
to a photon energy of 25 eV and convoluted with a
Gaussian having a profile width of 15 meV. Here again
strong $3s^23p^3(^2D^o_{5/2 })nd$ and $3s^23p^3(^2P^o_{1/2 })nd$ Rydberg
resonance series and the much weaker $3s^23p^3(^2D^o_{5/2 })ns$ and
$3s^23p^3(^2P^o_{1/2 })ns$ series are found in the respective spectra. 

The energies and quantum defects of all the Rydberg series
are tabulated in Table \ref{res-tab3} for both  series converging 
to the Cl$^{2+}$($^2P^o_{1/2}$) and Cl$^{2+}$($^2D^o_{5/2}$) thresholds 
for each of the respective $3s^23p^4\; ^1D_2$ and $^1S_0$ initial
states. The available ALS experimental results are included in Table \ref{res-tab3} for comparison purposes. 
From Table \ref{res-tab3}, for the $3s^23p^3nd$ Rydberg  series, we see
excellent agreement between the theoretical values and
the available ALS experimental measurements. It is seen
that the theoretical values for the resonance energies are
 within the $\pm$13 meV energy uncertainity of experiment. 
This provides additional confidence in our
cross sections results which are suitable for many
applications in astronomy and astrophysics 
and for laboratory plasma physics applications.

\section{Theory and experiment comparison}
To simulate experiment the theoretical photoionisation cross sections for each initial state was convoluted at the 
experimental resolution of 15 meV and a suitable admixture of the initial states present in the beam carried out.

Fig. \ref{overview} illustrates the comparison of our DARC calculations on the ground and metastable states of this sulphur-
like chlorine ion with the recent measurements from the ALS \citep{Hernandez2015}. In Fig. \ref{overview} 
we have convoluted the theoretical cross sections with a Gaussian having a profile 
width of 15 meV FWHM in order to compare with the experimental measurements. 
 To achieve the best match of the high resolution 
measurements made at the ALS \citep{Hernandez2015} with the DARC results we find that a non-statistical
weighting of the theoretical data for initial states of this sulphur-like chlorine ion 
was as follows; 1\% of the $3s^23p^4\; ^1S_0$
state, 27\% of the $3s^23p^4\; ^1D_2$ state, the other 
72\% distributed among the $3s^23p^4\; ^3P_{2,1,0}$ fine-structure levels, namely;
40\%  $^3P_{2}$, 24\%  $^3P_{1}$ and 8\% $^3P_{0}$. This weighting
of the initial states of the theoretical photoionisation cross sections appears 
to accurately reproduce the spectra in the high resolution ALS
measurements. 

An additional check on the theoretical data is the comparison of 
the integrated continuum oscillator strength $f$ with experiment. 
The integrated continuum oscillator strength $f$ of the experimental spectra was 
calculated over the energy grid [$E_1$, $E_2$], where $E_1$ is the minimum experimental energy  
and $E_2$ is the maximum experimental energy measured, (respectively 19.5 eV and 28 eV),  
using  \citep{Shore1967,Fano1968,berko1979},
\begin{equation}
f = 9.1075 \times 10^{-3} \int_{E_1}^{E_2} \sigma (h\nu) dh\nu
\end{equation}
and yielded  a value of 0.473 $\pm$ 0.095 from the ALS measurements. 
 A similar procedure for the theoretical {\it R}-matrix cross section (from appropriated weighted initial states) gave a
value of 0.405, in good agreement with experiment. This allows one to quantify an error estimate which we 
conservatively give as 15\% for the theoretical cross sections.

\section{Conclusions}

Large-scale Dirac Coulomb {\it R}-matrix (DARC) photoionisation cross section 
calculations have been carried for the singly ionised chlorine ion
for the $3s^23p^4\; ^3P_{2,1,0}$, $3s^23p^4\; ^1D_2$ and $3s^23p^4\; ^1S_0$ initial 
states in the energy range 19 - 28 eV.  For a physical understanding of the atomic processes 
taking place, we analyze the autoionizing resonance features found in the spectra. 
 Strong Rydberg series of the type $3s^23p^3nd$ along with
much weaker series of the form $3s^23p^3ns$ are found in the respective photoionisation spectra. All these resonance
series are analyzed, compared and contrasted with recently reported high resolution measurements made at
the ALS and spectroscoptically assigned.  Excellent agreement of the present theoretical photoionisation
cross sections with recent ALS measurements \citep{Hernandez2015} is found. 
This essential benchmarking of the theoretical work against high-resolution
 measurements provides confidence in the data for applications.

\section{Summary}

The photoionisation cross sections for the singly 
ionized atomic chlorine ion in the states $3s^23p^4\;^3P_{2,1,0}$, and
 $3s^23p^4\; ^1D_2$ and $3s^23p^4\; ^1S_0$ were calculated using the Dirac Coulomb 
{\it R}-matrix (DARC) method in the experimentally studied energy region of 19 - 28 eV. Numerous 
autoionization resonances features are found in the photoionisation spectra and their resonance energies and
quantum defects tabulated and compared with recent ALS measurements  \citep{Hernandez2015}. 
Various interloping resonances are found in the photoionisation spectra from the
different initial states of this sulphur-like species. 
The interloping resonances are seen to disrupt the regular Rydberg series pattern in the spectra. Overall the resonance
features found in the theoretical spectra obtained using the DARC codes in the valence photon energy 
region show  excellent agreement with previous ALS measurements \citep{Hernandez2015}
 and are within the $\pm$13 meV energy uncertainity of experiment.

The high resolution experimental measurements made at the ALS synchrotron radiation 
facility (over a limited energy range) have been used to benchmark the theoretical calculations,
and as such our theoretical work would be suitable to be incorporated into astrophysical modelling codes like 
 CLOUDY  \citep{Ferland1998,Ferland2003}, XSTAR \citep{Kallman2001} 
 and AtomDB  \citep{Foster2012} used to numerically
simulate the thermal and ionization structure of ionized astrophysical nebulae. 
All of the photoionisation cross sections are available (via email) from the author.

\section*{Acknowledgements}
BMMcL acknowledges support by the US National Science Foundation under the visitors program through a grant to ITAMP
at the Harvard-Smithsonian Center for Astrophysics and Queen's University Belfast through a visiting research fellowship (VRF).
BMMcL would like to thank Dr. Wayne Stolte and Dr. Alfred Schlachter for their hospitality during numerous visits 
to the Advanced Light Source in Berkeley, CA, where this work was completed.  
Professor Phillip Stancil is thanked for a careful reading 
of this manuscript and for many helpful suggestions.  
This research used resources of the National Energy Research Scientific Computing Center, 
which is supported by the Office of Science of the U.S. Department 
of Energy  (DOE) under Contract No. DE-AC02-05CH11231. 
The computational work was performed at the National Energy Research Scientific
Computing Center in Oakland, CA, USA 
and at The High Performance Computing Center Stuttgart (HLRS) 
of the University of Stuttgart, Stuttgart, Germany. 
This research also used resources of the Oak Ridge Leadership Computing Facility 
at the Oak Ridge National Laboratory, which is supported by the Office of Science 
of the U.S. Department of Energy (DoE) under Contract No. DE-AC05-00OR22725.

%
\bibliographystyle{mnras}
\bibliography{clplus}
%


\bsp	
\label{lastpage}
\end{document}